\documentclass[12pt]{article}
\usepackage{amssymb}
\usepackage{latexsym}
\usepackage[dvips]{graphicx}
\def\ga{\mathrel{\raise.3ex\hbox{$>$\kern-.75em\lower1ex\hbox{$\sim$}}}}
\def\la{\mathrel{\raise.3ex\hbox{$<$\kern-.75em\lower1ex\hbox{$\sim$}}}}
\def\beq{\begin{equation}}
\def\eeq{\end{equation}}

\def\fluxunit{cm$^{-2}$s$^{-1}$MeV$^{-1}$}
\def\sfrunit{M$_\odot$yr$^{-1}$Mpc$^{-3}$}
\def\snrate{yr$^{-1}$Mpc$^{-3}$}
\begin{document}
\begin{titlepage}
\pagestyle{empty}
\baselineskip=21pt
\rightline{\tt astro-ph/0509404}
\rightline{UMN--TH--2412/05}
\rightline{FTPI--MINN--05/40}
\rightline{September 2005}
\vskip 0.2in
\begin{center}
{\large{\bf Neutrino Signatures from the First Stars}}
\begin{center}
\end{center}
\vskip 0.2in
{\bf Fr\'ed\'eric~Daigne}$^1$, {\bf Keith~A.~Olive}$^{2}$, {\bf Pearl Sandick}$^3$,
and {\bf Elisabeth Vangioni}$^1$
{\it
$^1${Institut d'Astrophysique de Paris, UMR 7095, CNRS, Universit\'e Pierre et Marie Curie-Paris VI, 98 bis bd Arago, F-75014, Paris, France}\\
$^2${William I. Fine Theoretical Physics Institute, School of Physics and Astronomy,
University of Minnesota, Minneapolis, MN 55455 USA}\\
$^3${Department of Physics, School of Physics and Astronomy,
University of Minnesota, Minneapolis, MN 55455 USA}}

\vskip 0.1in
{\bf Abstract}
\end{center}
\baselineskip=18pt \noindent

Evidence from the WMAP polarization data indicates that the Universe may
have been reionized at very high redshift.   It is often suggested that
the ionizing UV flux originates from an early population of massive or very
massive stars.
Depending on their mass, such stars can explode either as
type II supernovae or pair-instability supernovae, or may entirely collapse into
a black hole. The resulting neutrino emission can be quite different in each case.
We consider
here the relic neutrino background produced by an early burst of Population III
stars coupled with a normal mode of star formation at lower redshift.
The computation is performed in the framework of 
hierarchical structure formation 
and is based
on cosmic star formation histories constrained to reproduce the observed star formation
rate at redshift $z \la 6$, the observed chemical abundances in damped Lyman alpha absorbers
and in the intergalactic medium, and to allow for an early reionization of the Universe
 at $z\sim 10-20$.
We find that although the high redshift burst of Population III stars does
lead to an appreciable flux of neutrinos at relatively low energy ($E_\nu \approx 1$ MeV),  the observable neutrino flux is dominated by the normal mode of star formation.  We also find that predicted
fluxes are at the present level of the SuperK limit.  As a consequence, the supernova relic neutrino
background has a direct impact on models of chemical evolution and/or supernova dynamics.

\end{titlepage}

\section{Introduction}

One of the most surprising results contained in the first year data obtained by WMAP \cite{wmap}
was large optical depth implying  that the Universe became reionized at high redshift in the range, $11<z<30$ at 95\% CL. This is much higher than previously believed, and it has been proposed that a generation of very massive stars is necessarily responsible~\cite{reion}. Even a brief period of massive star formation at high redshift would have notable consequences for chemical evolution, in particular,  the metal enrichment of the interstellar medium (ISM) and intergalactic medium (IGM) \cite{enrich,daigne1,daigne2}.  When combined with the observed cosmic star formation 
rate (SFR) at $z \la 6$ \cite{csfr}, a coherent picture of the 
star formation history of the Universe unfolds:  the first stars (Population III) 
described by a top-heavy initial mass function (IMF) were formed in primordial, 
metal-free structures with masses of order $10^7$M$_\odot$.
Once the metallicity achieved a certain critical level (of order $10^{-4}$ times the solar metallicity \cite{zcrit}), the massive mode of star formation yielded to a more normal distribution of stellar masses
at a rate over an order of magnitude larger than the current star formation rate. The cosmic SFR is 
observed to have peaked at redshift $z \approx 3$.

Among the consequences of this newly emerging view of star formation is the predicted enhancement
in the rate of core collapse supernovae.  In addition to the sharp spike of supernovae at very high redshift due to the explosions of stars responsible for the early epoch of reionization, the enhanced 
SFR of the normal mode of star formation at redshifts $z \la 6$ leads to a supernova rate
which is approximately a factor of 30 times the current rate, and a factor of 5 times the
observed rate at $z\sim 0.7$ \cite{dahlen}.

Another consequence of an enhanced SFR and SN rate is the resultant neutrino background spectrum
produced by the accumulated core collapse supernovae.  Early estimates of the (anti-e)-neutrino flux
based on simple models of Galactic chemical evolution were in the range of 1 -- 10 \fluxunit \cite{tot},
as were later estimates \cite{ando,as,sksw} based on peak SFRs of order 0.1 \sfrunit 
corresponding to supernova rates of order $10^{-3}$ \snrate.
In \cite{immrs}, the first attempt at
obtaining the neutrino flux from very massive Population III stars found fluxes at levels exceeding
10 \fluxunit, though the peak occurs at lower energy due to the redshifted spectrum.
Current estimates of the SFR peak at values 3 times higher at a redshift $z \sim 3$ \cite{csfr}
with correspondingly higher SN rates. More startling is the possibility that SN rates at high
redshift ($z \sim 15$) may be as large as $6 \times 10^{-2}$ \snrate.
Here, we incorporate fully developed chemical evolution models which trace the history of
pre-galactic structures as well as the IGM and are based on a $\Lambda$CDM cosmology and
include a Press-Schecter model \cite{ps} of hierarchical structure formation.
We adopt the approach and chemical evolution models of Daigne et al. \cite{daigne1,daigne2} and consider several bimodal IMFs, each with a normal component of star formation as well as  a 
massive component describing Population III stars. Given an IMF and a respective SFR, 
one can calculate the chemical history of cosmic structures and the IGM, the reionization efficiency, 
and as we do here,  the expected supernova relic neutrino (SRN) flux.

In section~\ref{sec:SRN} we describe our method for calculating the SRN flux, and in section~\ref{sec:SFmodels} we discuss the star formation models considered. In section~\ref{sec:results} 
we present our results for the SRN flux in a variety of models describing Population III stars. 
Opportunities for SRN detection are examined in section~\ref{sec:det}, and the effects of neutrino oscillations are discussed in section~\ref{sec:osc}. Our conclusions are given in section~\ref{sec:conc}.

\section{Supernova Relic Neutrino Background}\label{sec:SRN}

In all of the models we will consider, star formation begins at high redshift, dominated 
initially by massive stars which may explode as core collapse or pair-instability supernovae and
provide for the reionization of the IGM. Each explosion, regardless of type, releases most
of the star's gravitational energy in the form of neutrinos with a specified energy spectrum and flux.
Given a chemical evolutionary model, or more specifically, given a rate of star formation along
with the IMF, the integrated contribution of SN to the neutrino background can be computed.

The expected differential flux of neutrinos at Earth with energy $E$ can be expressed as 
\beq
\frac{dF_\alpha}{dE} =
\int_{0}^{z_i} dz\, (1+z) \left| \frac{dt}{dz} \right|
\int_{M_{min}}^{M_{max}} dm\, \phi(m) \, \psi\left(t-\tau(m)\right) \,
N_{\nu_\alpha}(m)\, \frac{dP_\alpha}{dE'}
\label{diffnu}
\eeq
where $N_{\nu_\alpha}(m)$ is the total number of neutrinos of a given species, $\alpha$, emitted in the core collapse of a star of mass $m$, $\phi$ is the normalized IMF, $\psi$ is the SFR per unit comoving volume, $\tau(m)$ is the lifetime of a star of mass $m$,  and $\frac{dP_\alpha}{dE'}$ represents the neutrino spectra in the comoving volume at  energy $E' = E(1+z)$. The integration limits $M_{min}$ and $M_{max}$ are the minimum and maximum masses in each model for which supernovae occur, and $z_i$ is the initial  redshift over which star formation occurs. It is assumed that star formation continues to the present ($z = 0$).

When a star undergoes core collapse, the mass of the remnant is determined by the mass of the progenitor.  We assume that all stars of mass $ m \gtrsim 8$ M$_{\odot} $ will die as supernovae. For stars of mass 8 M$_{\odot} < m < 30$ M$_{\odot} $, the remnant after core collapse will be a neutron star of $ m \approx 1.5$~M$_{\odot} $. More massive stars fall into two categories; black holes and pair instability supernovae. Pair instability supernovae are thought to occur for stars with 140~M$_{\odot} \lesssim m \lesssim 260$ M$_{\odot} $, in which case the explosion leaves no remnant. All other stars collapse to form black holes. Stars with 30 M$_{\odot}<m<100$ M$_{\odot} $ become black holes with mass approximately that of the star's helium core before collapse~\cite{stardeath}. We take the mass of the Helium core to be 
\beq
M_{He}=\frac{13}{24} \cdot (m-20\,M_{\odot}) 
\eeq
for a star with main sequence mass $m$~\cite{ww2002}. We assume that stars with $ m>260$ M$_{\odot} $ collapse entirely to black holes.

The energy emitted in each core collapse, $E_{cc}$ corresponds to the change in gravitational energy, 99\% of which is emitted as neutrinos~\cite{by}. In the cases where collapse results in a neutron star, $E_{cc}=5 \times 10^{53}\,$ergs. For stars that collapse to black holes, $E_{cc}$ is proportional to the mass of the black hole. Pair instability supernovae experience a much more powerful explosion than core collapse supernovae, however few neutrinos are emitted and with very low energies such that they would not be observed~\cite{wwm1986}. We address this issue further in section~\ref{sec:SFmodels}. Although several studies find a distinct hierarchy in the partitioning of neutrino luminosity among the species during the different luminosity phases of core collapse, equipartition of the total energy emitted by the star is generally accepted~\cite{tot,as,sksw}. For a comparison of luminosity hierarchies found in recent simulations, see Keil et al.~\cite{krj}. 

In black hole formation, the neutrino luminosity is nearly constant for the first few seconds until the event horizon overtakes the neutrinospheres. Once the neutrinospheres are inside the event horizon, the luminosity consists of neutrinos with lower average energies escaping from the outer layers of the star~\cite{fwh, krj}. We assume that each electron neutrino carries an average energy $\langle E_{\nu_e}\rangle = 13.3 \,\textrm{MeV}$, which is a reasonable approximation for the average neutrino energy over the two luminosity phases of core collapse to a black hole. For supernovae which do not collapse to black holes, this energy is consistent with recent simulation data~\cite{krj}. The charged current reactions that prevent neutrinos from emerging from the star are $\nu
_{\textrm{e}} \textrm{n} \to \textrm{p} \textrm{e}^-$ and $\bar{\nu}_\textrm{e}\textrm{p} \to \textrm{n} \textrm{e}^+$. The different trapping reactions result in different neutrinosphere radii, and therefore different average energies for $\nu_e$ and $\bar{\nu}_e$. We assume $\langle E_{\bar{\nu}_e}\rangle = 15.3 \,\textrm{MeV}$ which is the average energy over the two luminosity phases as above, following~\cite{immrs}. The other species, denoted $\nu_x$, undergo only neutral current interactions.  The mechanism that governs their average temperature at emission is more complicated, but the generally accepted hierarchy is $\langle E_{\nu_e}\rangle < \langle E_{\bar{\nu}_e}\rangle < \langle E_{\nu_x}\rangle$. 
We have taken $\langle E_{\nu_x} \rangle = 20$ MeV.
The total number of $\nu_{\alpha}$ emitted by a star during core collapse is given by 
\beq
N_{\nu_\alpha}=\frac{E_{cc}}{\langle E_{\nu_\alpha}\rangle}.
\eeq
We discuss the sensitivity of our results to these choices in section~\ref{sec:det} below.

The neutrino spectra can be described by a normalized Fermi-Dirac distribution, 
 \beq
  \frac{dP_{\alpha}}{dE'}=\frac{2}{3 \zeta_3 T_{\alpha}^3} \frac{{E'}^2}{e^{E'/T_{\alpha}}+1}
\eeq
 where $T_{\alpha}= {180 \zeta_3 \langle E_{\nu_\alpha}\rangle}/{7 \pi^4}$ is the effective neutrino temperature taken to be independent of the mass of the star.
 We assume a flat $\Lambda$CDM cosmology with  
 \beq
 \left| \frac{dt}{dz} \right| = \frac{9.78\,h^{-1}\,\textrm{Gyr}}{(1+z)\sqrt{\Omega_{\Lambda}+\Omega_m(1+z)^3}}
 \eeq
 where $\Omega_{\Lambda}=0.73$, $\Omega_m=0.27$, and $h=0.71$~\cite{wmap}. 

\section{Star Formation Models}\label{sec:SFmodels}

The cosmic star formation histories we consider here have been adopted from detailed chemical
evolution models \cite{daigne1,daigne2}.  These models are bimodal and are described by a 
birthrate function of the form
\begin{equation}
B(m,t) = \phi_1(m) \psi_1(t) + \phi_2(m) \psi_2(t)
\label{birth}
\end{equation}
where $\phi_{1(2)}$ is the IMF of the normal (massive) component of star formation, and 
$\psi_{1(2)}$ is the respective star formation rate. 
The normal component contains stars with masses between 0.1 and 100 M$_\odot$
and is primarily constrained by observations at low redshift ($z \la 6$). The massive component operates  at high redshift and is required by the evidence for the reionization of the Universe at $z \sim 17$. 
Both components can contribute to the chemical enrichment of galaxy forming structures and the IGM.
We consider three different models of the massive mode as described below.

Given an IMF and SFR, it is straightforward to compute the
rate of core collapse supernovae,
\begin{equation}
{\mathrm SNR} =\int_{max(8\mathrm{M_{\odot}}, m_\mathrm{min}(t))}^{m_\mathrm{sup}} 
dm\ \phi(m)\psi\left(t-\tau(m)\right) \ ,
\end{equation}
where $m_\mathrm{min}(t)$ is the minimum mass with lifetime less than $t$.
Then for each model, the SRN flux is calculated using eq. (\ref{diffnu}).
 In the two following subsections,
we describe the two evolution models we have used, and their corresponding SFR and IMF.

\subsection{Stage 1: Simplified Models}
\label{sec:SFmodels:calc}

We begin by describing a simplified set of hierarchical models \cite{daigne1}. 
We assume star formation to occur between $z_i=20$ and $z_f=0$. 
The  IMF for each mode is similar to a Salpeter mass function, 
\beq 
\phi_i (m) \propto m^{-(1+x) }
\eeq
 but has a slightly steeper slope of $x=1.7$.
 Each IMF is normalized independently by
 \begin{equation}
\int_{m_\mathrm{inf}}^{m_\mathrm{sup}} dm\ m \phi_i(m)=1\ ,
\label{norm}
\end{equation}
differing only in the specific mass range of each model.

The mass in cosmic structures will be denoted $M_{struct}(t)$, which includes both the mass in stars and the mass in gas of  the ISM. 
 The mass in the IGM is $M_{IGM}$ and the total mass $M_{tot}=M_{IGM}+M_{struct}$ is
of course constant.
The mass in structures evolves as 
\beq 
\frac{dM_{struct}}{dt}=a_b(t)-o(t)\, ,
\label{gr0}
\eeq 
where $a_b(t)$ is the cosmic baryon accretion rate 
due to the structure formation process and $o(t)$ is the outflow from structures. 
In our simplified set of models, which we denote as stage 1, 
we will ignore the effects of outflow and set $o(t) = 0$.
We will also assume structure formation to be exponentially decreasing from time $t=0$, 
\beq
a_b(t)=\frac{a}{\tau_s}M_{tot}e^{-t/\tau_s}\, ,
\label{ab0}
\eeq
where $a = 0.1$ is the fraction of the total mass which is eventually accreted by structures and $\tau_s$ is the timescale of the accretion process. Here we consider both $\tau_s=0.01 \,\textrm{Gyr}$ and $\tau_s=0.2 \,\textrm{Gyr}$~\cite{daigne1}. We also assume that star formation begins when the baryon fraction in structures is  $1\%$.  This is our initial condition at $z=20$ and corresponds to an estimate of the minimum baryon fraction where sufficient dissipation occurs to allow star formation \cite{silk}.  In our more complex models 
we retain the condition on the onset of star formation, but determine its redshift using a detailed model of
hierarchical structure formation (see next subsection). 

The evolution of the gas mass in the structures is given by\,:
\begin{equation}
\frac{d M_\mathrm{ISM}}{dt} = -\psi(t) + e(t) +
a_\mathrm{b}(t) - o(t)\ .
\label{eq:dmismdt}
\end{equation}
The stellar mass in the structures is simply $M_\mathrm{struct}-M_\mathrm{ISM}$.  
Each term in equation~(\ref{eq:dmismdt}) accounts for part of the gas budget of the ISM.  
The first term corresponds to the loss of gas through star formation while the
second term corresponds to the ejected gas when the star dies.
Here, we will use the instantaneous recycling approximation (IRA) to evaluate the
 rate at which gas is returned to the ISM, $e(t)$
 \begin{equation}
e(t) = \int_{0.9 \mathrm{M}_\odot}^{m_{sup}}dm\ \phi (m)\psi (t) (m-m_\mathrm{r})
\equiv R \psi(t)\ .
\label{eq:e}
\end{equation} 
where $m_\mathrm{r}$ is the mass of the remnant, and $R$ is the IRA return mass fraction.

The normal mode of star formation in \cite{daigne1} is referred to as Model 0 and provides a standard star formation history, with stellar masses in the range 0.1 M$_{\odot} \le m \le 100$ M$_{\odot}$. The SFR for Model 0 is proportional to the gas mass fraction in cosmic structures, $\sigma = M_{ISM}/M_{struct}$,
\beq
\psi_1 = \nu_1 \sigma(t)\, ,
\eeq 
where $\nu_1=M_{struct}(t)/\tau_1$ with $\tau_1=5\,\textrm{Gyr}$ which is a typical timescale for star formation in the galactic disk. This model alone is inadequate for high redshift reionization.

We consider three different models, labeled Models 1, 2a, and 2b to describe the massive mode.
They are distinguished by their respective stellar mass ranges.
In Model 1,  the IMF is defined for stars with masses, 40 M$_{\odot} \le m \le 100$ M$_{\odot}$. 
All of these stars die in core collapse supernovae leaving a black hole remnant.
They all contribute to the chemical enrichment of the ISM (and IGM when $o(t) \ne 0$).
This period of star formation is brief and is described by a SFR of 
the form
\beq
\psi_2=\nu_2e^{-t/\tau_2}
\eeq 
where $\nu_2=f_2\,M_{struct}(t)/\tau_2$ with a characteristic timescale
$\tau_2=50\,\textrm{Myr}$. The constant $f_2=4.5\%$.

Model 2a is described by
very massive stars which become pair instability supernovae.  The IMF is defined for 140 M$_{\odot} \le m \le 260$ M$_{\odot}$ and the SFR for this model is the same as that for Model 1, but must be reduced by a factor of 8 due to constraints on metal abundances in the ISM. We consider the best case scenario for observation, where the energy emitted in neutrinos is the same as that for ordinary core collapse supernovae, but the average neutrino energy is $\langle E_{\bar{\nu}_e} \rangle = 1.2 \, \textrm{MeV} $~\cite{wwm1986}.

The most massive stars are considered in Model 2b and fall in the range 270 M$_{\odot} \le m \le 500$ M$_{\odot}$, with the SFR as in Model 1. These stars entirely collapse into black holes and do not contribute 
to the chemical enrichment of either the ISM or IGM.  Collectively we will refer to stage 1 models 
as 1.0, 1.1, 1.2a, and 1.2b. Unless otherwise noted, Models 1.1, 1.2a, and 1.2b (as well 2.1, 2.2a, and 2.2b defined below) will correspond to the full bimodal model, i.e., they include the normal mode
of star formation as well as the particular massive mode.

\subsection{Stage 2: Hierarchical Models}
\label{sec:SFmodels2:calc}

We also consider a set of models with a more 
sophisticated treatment of the hierarchical growth of structure which we will call here stage 2. 
Complete details of this model can be found in \cite{daigne2}.
Rather than assuming an analytic form for the formation of structure as in eq. (\ref{ab0}),
we use the Press-Schechter formalism and take 
\begin{eqnarray}
a_\mathrm{b}(t) & = & \Omega_\mathrm{b}\left(\frac{3H_{0}^{2}}{8\pi G}\right)\ \left(\frac{dt}{dz}\right)^{-1}\ \left|\frac{d f_\mathrm{b,struct}}{dz}\right|\nonumber\\
&  \approx  & 1.2 h^{3}\ \mathrm{M_{\odot}/yr/Mpc^{3}}\ \left(\frac{\Omega_\mathrm{b}}{0.044}\right)\left(1+z\right)\sqrt{\Omega_\mathrm{\Lambda}+\Omega_\mathrm{m}\left(1+z\right)^{3}}\left|\frac{df_\mathrm{b,struct}}{dz}\right|\ .
\end{eqnarray}
where $f_b$ is the baryon fraction in structures
\begin{equation}
f_\mathrm{b,struct}(z) = \frac{\int_{M_\mathrm{min}}^{\infty} dM\ M f_\mathrm{PS}(M,z)}{\int_{0}^{\infty} dM\ M f_\mathrm{PS}(M,z)}\ .
\label{eq:fb}
\end{equation}
In eq. (\ref{eq:fb}), $f_\mathrm{PS}(M,z)$ is the distribution function of halos and is computed using the method described in \cite{ps}, and $M_\mathrm{min}$ is the minimum halo mass in which stars are formed.  As before, we require an initial baryon fraction $f_b =  1$\%, which for example for $M_\mathrm{min} = 10^7$ M$_\odot$ corresponds to an initial redshift for star formation, $z = 16$.

In \cite{daigne2}, it was found that better fit to the global star formation rate and supernova
rates is obtained when a normal mode SFR of an exponential form
\beq
\psi_1=\nu_1e^{-t/\tau_1}
\eeq 
is used, which corresponds to a SFR dominated by elliptical galaxies.  
Best fits for $\nu_1$ and $\tau_1$ are given in Table 1.
In these models the outflow is non-zero and the details for computing the outflow are given in \cite{daigne1,daigne2}. The overall efficiency of outflow is parameterized by $\epsilon$ whose
value is also given in Table 1. The instantaneous recycling approximation is no longer used here,
and ejection rates depend on stellar lifetimes, $\tau(m)$.  This amounts to replacing
$\psi(t)$ with $\psi(t-\tau(m))$ in the integral in eq. (\ref{eq:e}).

\begin{table}[ht]
\begin{center}
\begin{tabular}{|ccccc|}
\multicolumn{5}{c}{\textbf{Normal mode}}  \\ 
\hline
$M_{\mathrm{min}}$ & 
$z_{\mathrm{init}}$ & 
$\epsilon$ & 
$\nu_{1}$  & 
$\tau_{1}$ \\
(M$_{\odot}$) & 
 & 
 & 
(Gyr$^{-1}$) & 
(Gyr)  \\
\hline
\hline
$10^{6}$  & 18.2 & $2\times 10^{-3}$ & 0.2 & 2.8 \\
\hline
$10^{7}$  & 16.0 & $3\times 10^{-3}$ & 0.2 & 2.8 \\
\hline
$10^{8}$  & 13.7 & $5\times 10^{-3}$ & 0.2 & 2.8 \\ 
\hline
$10^{9}$  & 11.3 & $        10^{-2}$ & 0.2 & 3.0 \\ 
\hline
$10^{11}$ & 6.57 & $1.5\times 10^{-2}$ & 0.5 & 2.2  \\ 
\hline
\end{tabular}
\end{center}
\caption{The model parameters for the normal mode of star formation (Model 2.0). Column 1 indicates the input value of the minimum mass for star forming structures. Column 2 is derived from column 1, having assumed that $f_b = 1\%$ when star formation begins.  In columns 3, 4, and 5, parameter values for the efficiency of outflow and the SFR are given.  The slope of the IMF is $x=1.3$ for all models. 
}
\label{tab:model0}
\end{table}

The SFRs for the massive modes in stage 2 are determined by the metallicity in the ISM,
\beq
\psi_2=\nu_2e^{-Z/Z_\mathrm{crit}}
\eeq 
where $Z_\mathrm{crit} = 10^{-4}$ Z$_\odot$ is the critical metallicity at which Population III star formation 
ends \cite{zcrit}.  
The IMF of both modes in this case
has a slope $x_{1(2)} = 1.3$.  
The massive mode SFR parameters are given in Table 2 for all models considered.
The mass ranges and neutrino average energies in stage 2 are the same as those in stage 1 respectively.

\begin{table}[ht]
\begin{center}
\begin{tabular}{|lccc|}
\multicolumn{4}{c}{\textbf{Massive mode}}  \\ 
\hline
Model & $M_{\mathrm{min}}$ & 
$\epsilon$ & 
$\nu_{2}$  \\
& (M$_{\odot}$) 
 & 
 & 
(Gyr$^{-1}$)  \\
\hline
\hline
2.1 & $10^{7}$  &  $2 \times 10^{-3}$ & 60 \\
\hline
2.1e & $10^{7}$  &  $6 \times 10^{-5}$ & 340 \\
\hline
2.2a & $10^{7}$  &  $1 \times 10^{-3}$ & 9 \\
\hline
2.2ae & $10^{7}$  &  $8 \times 10^{-5}$ & 40 \\
\hline
2.2b & $10^{7}$  &  $3 \times 10^{-3}$ & 100 \\
\hline
\end{tabular}
\end{center}
\caption{Parameter values for the massive starburst Models 2.1, 2.2a and 2.2b. Column 1 indicates the 
model number and column 2 the input value of the minimum mass for star forming structures.   In columns 3 and 4, we show the adopted outflow efficiency and massive mode SFR.}
\label{tab:model1}
\end{table}

When a massive mode is added to the normal mode described by Model 2.0,
the outflow efficiency must be adjusted so as to avoid the overproduction of 
metals in the IGM.  However, there is a degeneracy in the massive mode parameters
$\epsilon$ and $\nu_2$.  In models labeled 2.1 and 2.2a, the massive mode
contributes roughly 50 \% of the IGM metallicity at a redshift $z = 2.5$.
By increasing $\nu_2$ and decreasing $\epsilon$, this contribution can
be increased to 90 \% and at the same time increases the ionization capacity of the model.
These cases are labeled 2.1e and 2.2ae\footnote{Since stars associated with Model 2b do
not contribute to element enrichment, there is no Model 2be.}.

\section{Results}\label{sec:results}

\subsection{Neutrino Background}\label{sec:results:SRNbg}

Stage 1 and Stage 2 models are discussed individually in sections~\ref{sec:results:SRNbg:stage1} and~\ref{sec:results:SRNbg:stage2}. Then comparisons are made in section~\ref{sec:results:SRNbg:comp}
with other results.
We will for the most part consider only $\bar{\nu}_e$, as they are the most easily detected at water Cerenkov detectors. We will return to $\nu_e$ when we address thermonuclear neutrinos and detection.

\subsubsection{Stage 1}\label{sec:results:SRNbg:stage1}

\begin{figure}[ht]
\centering
\includegraphics[width=0.58\textwidth]{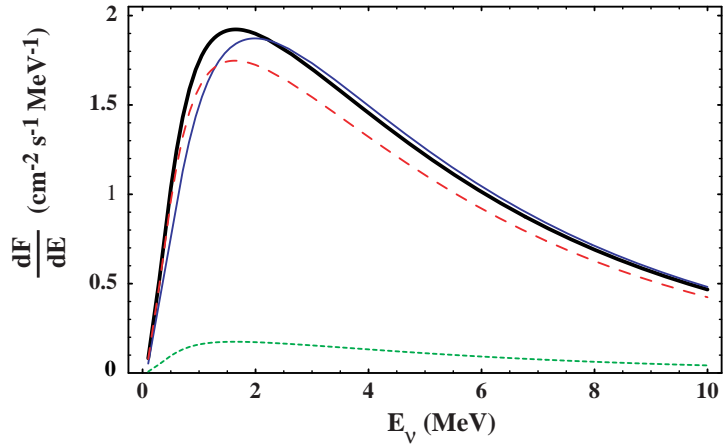}
\caption{Electron anti-neutrino fluxes from Model 1.0 with different baryon accretion rates. We have chosen two models with $a_b=0$ with different baryon fractions, $M_{struct}/M_{tot}=.1$ (red dashed) and $M_{struct}/M_{tot}=.01$ (green dotted).  We have also chosen two models with structure growth using $\tau_s=0.01\textrm{Gyr}$ (thick black), and  $\tau_s=0.2\,\textrm{Gyr}$ (solid blue). In the latter cases, $a = 0.1$.
\label{model0}}
\end{figure}

We begin by examining the neutrino production from simplified models, labeled here as stage 1.
As described above, these models employ the IRA and as such can to a large extent be
treated (semi)-analytically.  In this class of models, outflows are ignored and hierarchical growth
is simply modeled by eqs. (\ref{gr0}) and (\ref{ab0}).  All of the Population III models we consider are bimodal (see eq. (\ref{birth})) and combine a normal stellar distribution (model 0) along
with a massive mode. As a reference point, we first compute the expected neutrino flux from
the normal mode alone.  This is shown in Fig. \ref{model0}, where we show the flux of $\bar{\nu_e}$ only.  The peak flux is about 2 \fluxunit and occurs at $E_\nu \approx$ 2 MeV.  
While this flux is small compared to solar neutrino fluxes (for example,
the flux of $^8$B neutrinos from the sun is of order $10^6$~\fluxunit), it is large compared to the 
non-localized atmospheric neutrino flux which is less than $10^{-2}$ \fluxunit (see e.g. \cite{ando}). 

Fig. \ref{model0} shows the resultant neutrino flux for two choices of the baryon fraction in 
structures, $M_{struct}/M_{tot}=.1$ (large dashes) and for comparison, $M_{struct}/M_{tot}=.01$ (small dashes) when the growth of structures is neglected (i.e. $a_b = 0$).  As expected, the neutrino
flux is in direct proportion to the baryon fraction.   We also show in Fig. \ref{model0}, the neutrino
flux when $a_b \ne 0$, and the mass of the structure grows.  We show results for two different growth constants, $\tau_s = 0.01$ and 0.2 Gyr. Since the final baryon fraction is 10\% in each case, 
we see that the integrated neutrino flux is quite similar. 

 \begin{figure}[htb]
\centering
\includegraphics[width=0.58\textwidth]{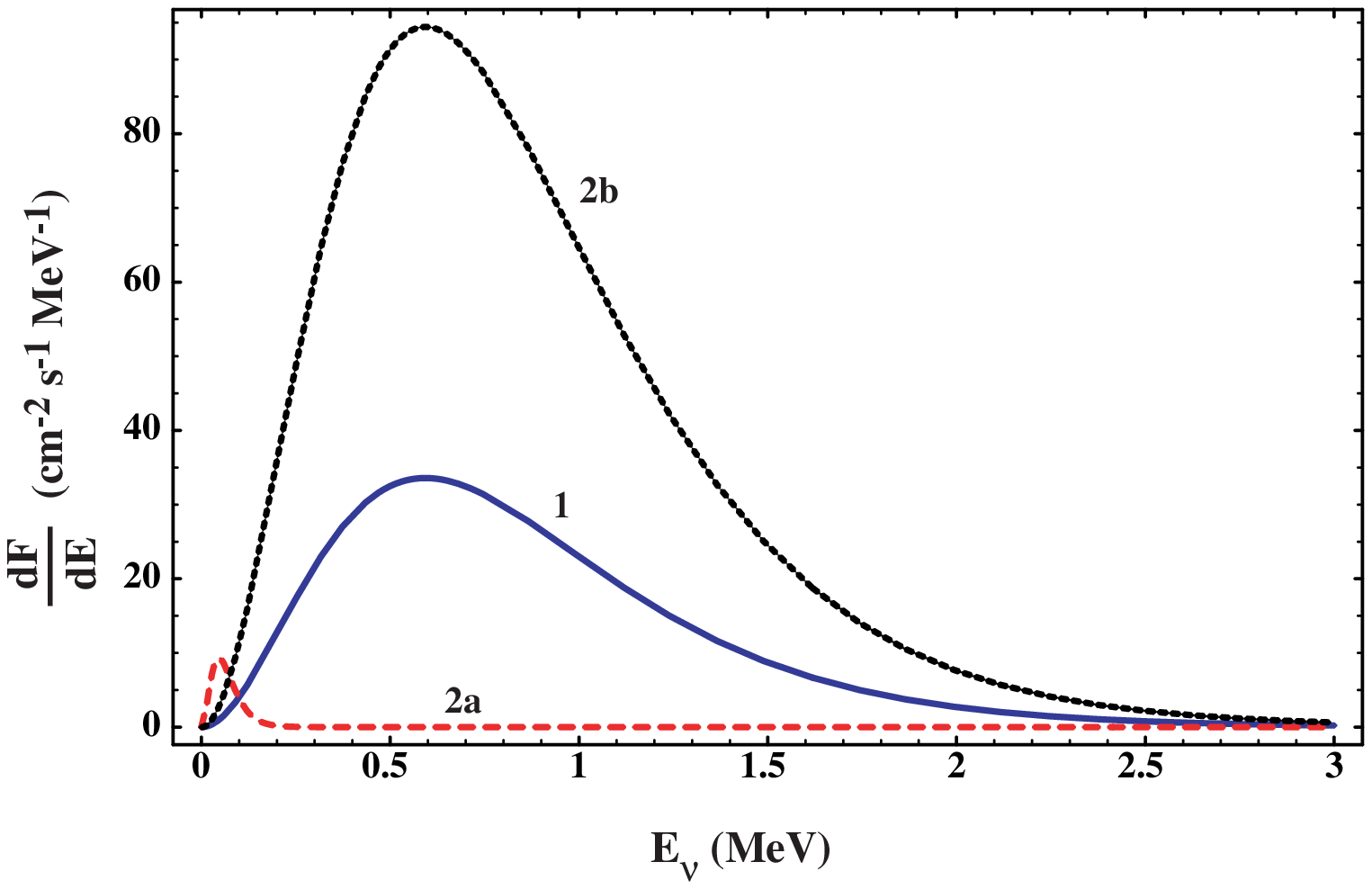}
\caption{Fluxes from  the massive modes of Models 1.1 (solid-blue), 1.2a (dashed-red), and 1.2b (dotted-black).
\label{models12}}
\end{figure}

The individual fluxes for the massive modes of Models 1.1, 1.2a, and 1.2b are displayed in Figure~\ref{models12}. For each of the massive modes, the duration of star formation is very brief,
as the SFR is characterized by a time constant $\tau_2 = 0.05$ Gyr.  The neutrino fluxes for 
Models 1.1 and 1.2b, shown by solid (blue) and dotted (black) curves respectively, are significantly 
larger than those found for a normal population of stars. In these cases, the fluxes are approximately 
30 and 95 \fluxunit, though the peak of the spectrum occurs at lower energy, $E_\nu \simeq 0.6$ MeV.
 We note that the SRN flux from Model 1.2b is very similar to the flux due to the population of rotating 300 M$_\odot$ stars with a SFR peaked at $z=17$ in Iocco et al.~\cite{immrs}. The peak height obtained in our calculations is larger due primarily to the fact that the integrated fraction of baryonic matter in population III stars is about four times greater in our model. It has been noted that if the collapse to a black hole proceeds without rotation, the neutrino luminosity will be diminished by $\sim 2$ orders of magnitude~\cite{fwh}, so this is really an upper limit to the flux assuming rotation.

 The flux from Model 1.2a is much smaller than the fluxes from Models 1.1 and 1.2b, and peaks at a lower energy. This is a consequence of less energy being released in neutrinos by pair instability supernovae than by supernovae that collapse to form black holes, and that the average energy of each neutrino is limited by silicon burning and photodisintegration~\cite{wwm1986}.

Comparison of Figs. \ref{model0} and \ref{models12} shows that the normal mode, though smaller 
in its peak flux, dominates the full bimodal spectrum at energies $E \ga 2.5$ MeV.  Recall that the massive mode is very localized at high redshift. As a result, neutrinos produced by the massive mode have energies which are redshifted  from their initial value.  In contrast, the normal mode of star formation is peaked at $z \simeq 2 - 3$, which produces a broader spectrum today.

\begin{figure}[ht]
\centering
\includegraphics[width=0.58\textwidth]{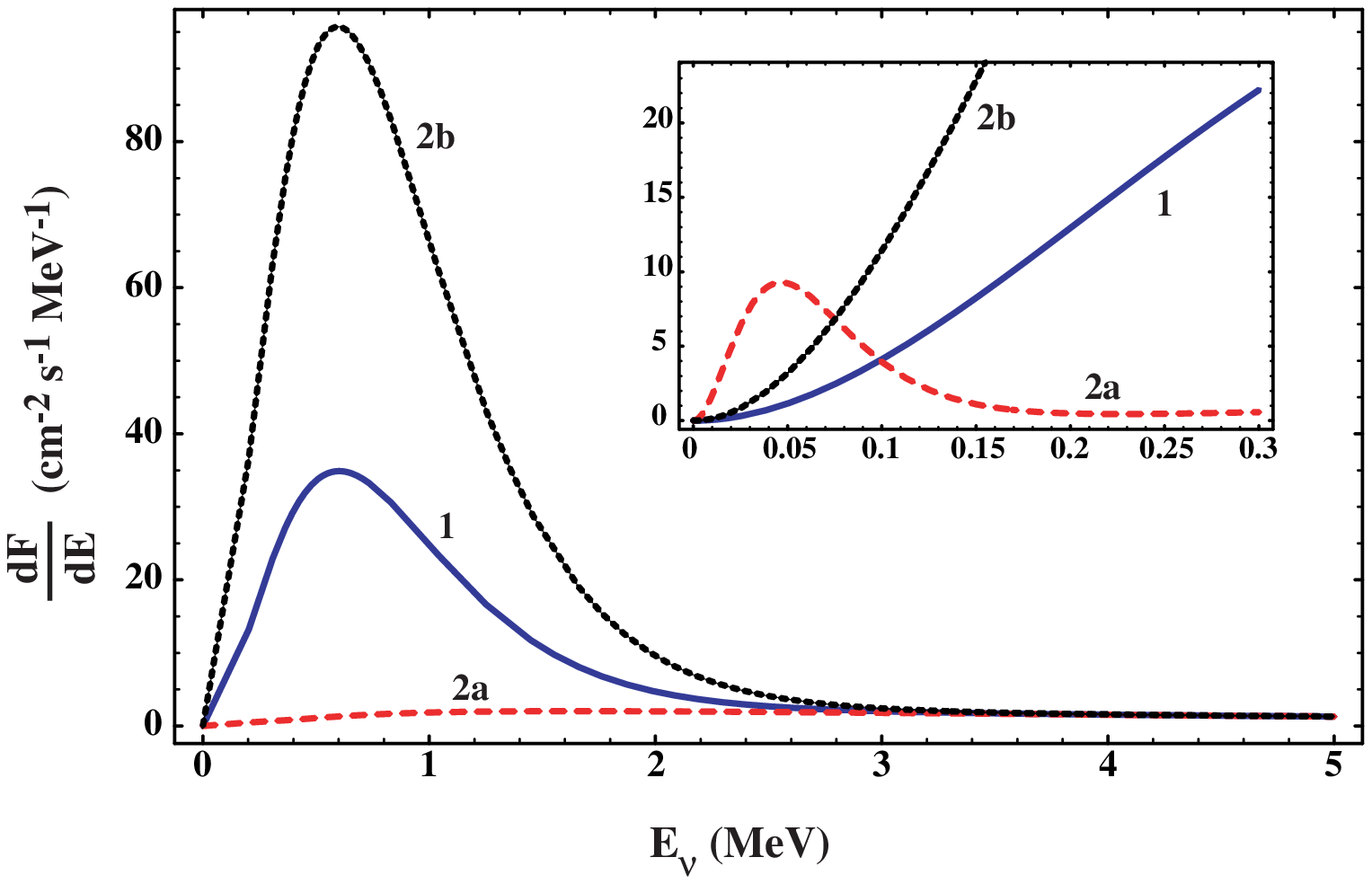}
\caption{Total fluxes from Models 1.1 (solid-blue), 1.2a (dashed-red), and 1.2b (dotted-black). In all three cases, $a_b \ne 0$ and $ {\tau_s=0.01\,\textrm{Gyr}}$. The insert shows the low energy peak due to PISN.
\label{totalflux}}
\end{figure}

Finally, Figure~\ref{totalflux} shows the total flux for the bimodal IMFs in the case that the massive component is either Model 1.1, 1.2a, or 1.2b. At low energies, the flux is dominated by the massive component.  The insert shows the low energy peak due to PISN.

\subsubsection{Stage 2}\label{sec:results:SRNbg:stage2}

\begin{figure}[ht]
\centering
\includegraphics[width=0.8\textwidth]{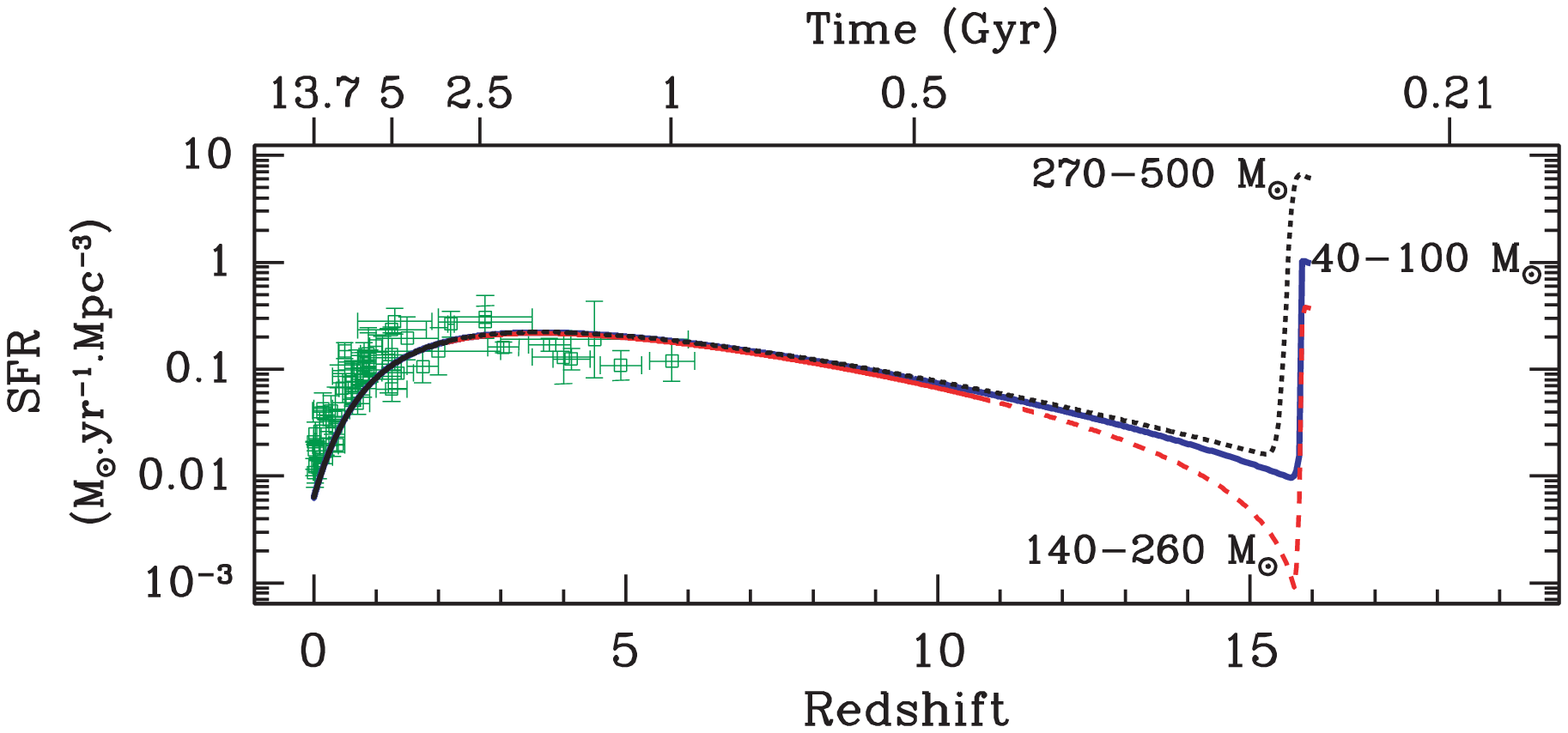}
\caption{The SFR for Model 2.1 (solid-blue), 2.2a (dashed-red), and 2.2b (dotted-black) as a function of redshift. Each model is indicated by the mass range associated with the massive mode.
In all three cases, $M_\mathrm{min} = 10^7$ M$_\odot$. Data are taken from \cite{csfr}.
\label{sfr}}
\end{figure}

We now use the full numerical results of the models described in \cite{daigne2}.
The overall form of these models is similar to those in stage 1, i.e. they are described by a bimodal 
birthrate function of the form in eq. (\ref{birth}).  The mass ranges for models 0, 1, 2a, and 2b
are unchanged, though the slope of the IMF in each case is now 1.3.  As described above,
these models include both outflow and the hierarchical growth of structure and we no longer 
employ the IRA in any of our calculations.  The SFR is shown in Fig. \ref{sfr}
for Models 2.1, 2.2a, and 2.2b as indicated by their respective mass ranges.
In Models 2.1 and 2.2a, the massive mode contributes roughly equally with the normal mode
to the IGM metallicity at $z \sim 3$.  Since Model 2.2b produces no heavy elements, 
the SFR is chosen to be maximal to enhance its ionization potential.
In each case, star formation begins at $z \sim 16$ when the baryon fraction in a structure of 
total mass $10^7$ M$_\odot$ is $f_b = $ 1\%. The massive burst ends when the ISM metallicity has reached a critical
value taken to be $10^{-4}$ Z$_\odot$.  The duration of the burst in Model 2.2b is somewhat
more prolonged as the metallicity is produced solely by the normal component.

\begin{figure}[ht]
\centering
\includegraphics[width=0.58\textwidth]{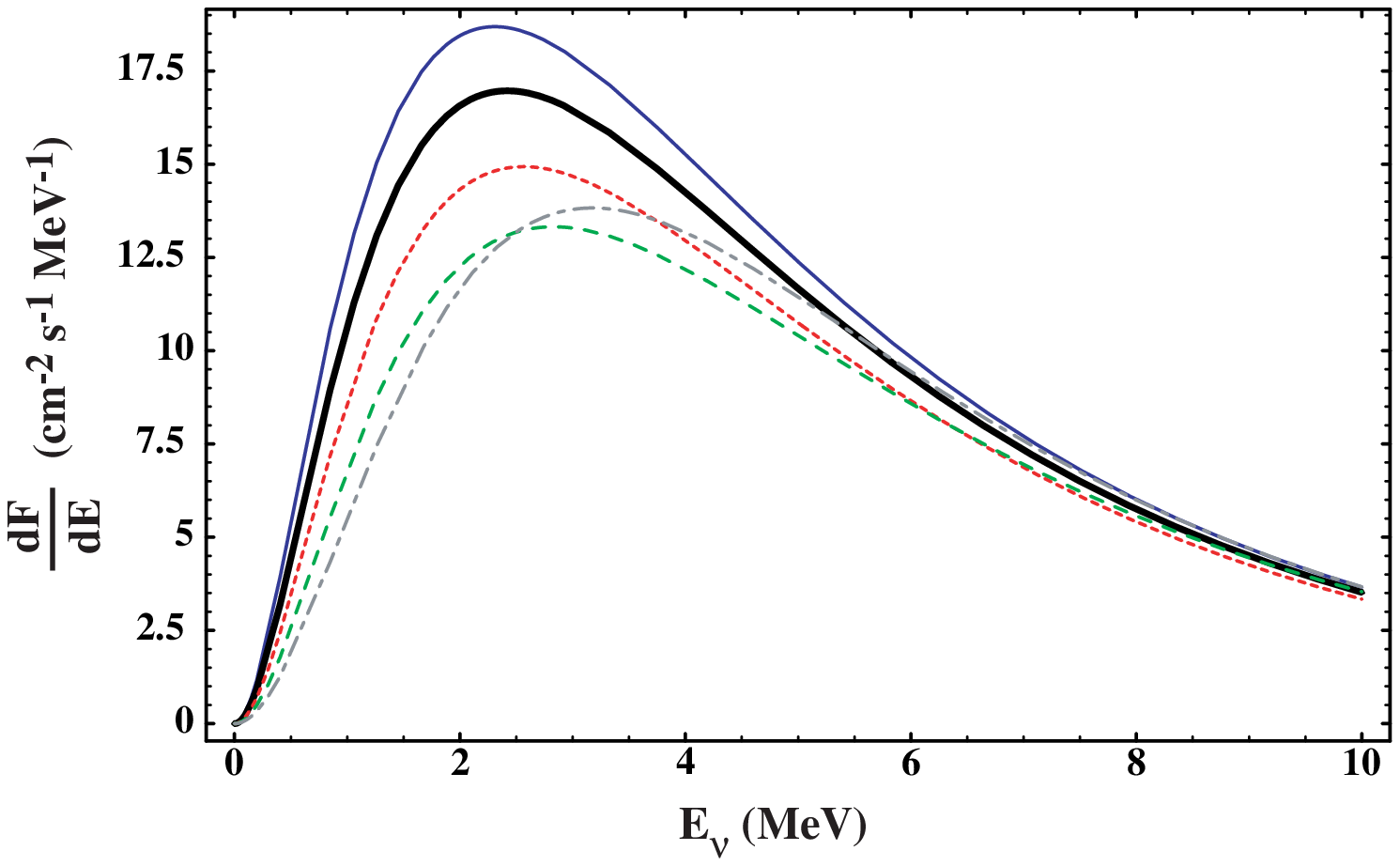}
\caption{Fluxes from Model 2.0 for five choices of $M_\mathrm{min} = 10^6$ (solid blue), $10^7$ (thick black), $10^8$ (dotted red), $10^9$ (dashed green) and 
$10^{11}$ (dot-dashed grey) M$_\odot$.}
\label{model0s2}
\end{figure}

Model 2.0 fluxes are plotted in Figure~\ref{model0s2}. Results are shown for several choices of minimum halo masses, $M_\mathrm{min}$. The neutrino fluxes found here are roughly a factor of 10 times larger than that found in
the semi-analytical model discussed in the previous subsection.  This is partly due to the 
flatter IMF chosen here which greatly enhances the numbers of massive stars and hence the supernova rate and neutrino flux.   
As we will see below these fluxes are large enough to be probed by
current detectors. As $M_\mathrm{min}$ is increased, star formation occurs at later redshift and as a 
result, the peak of the neutrino flux is shifted slightly to higher energy.

\begin{figure}[ht]
\centering
\includegraphics[width=0.58\textwidth]{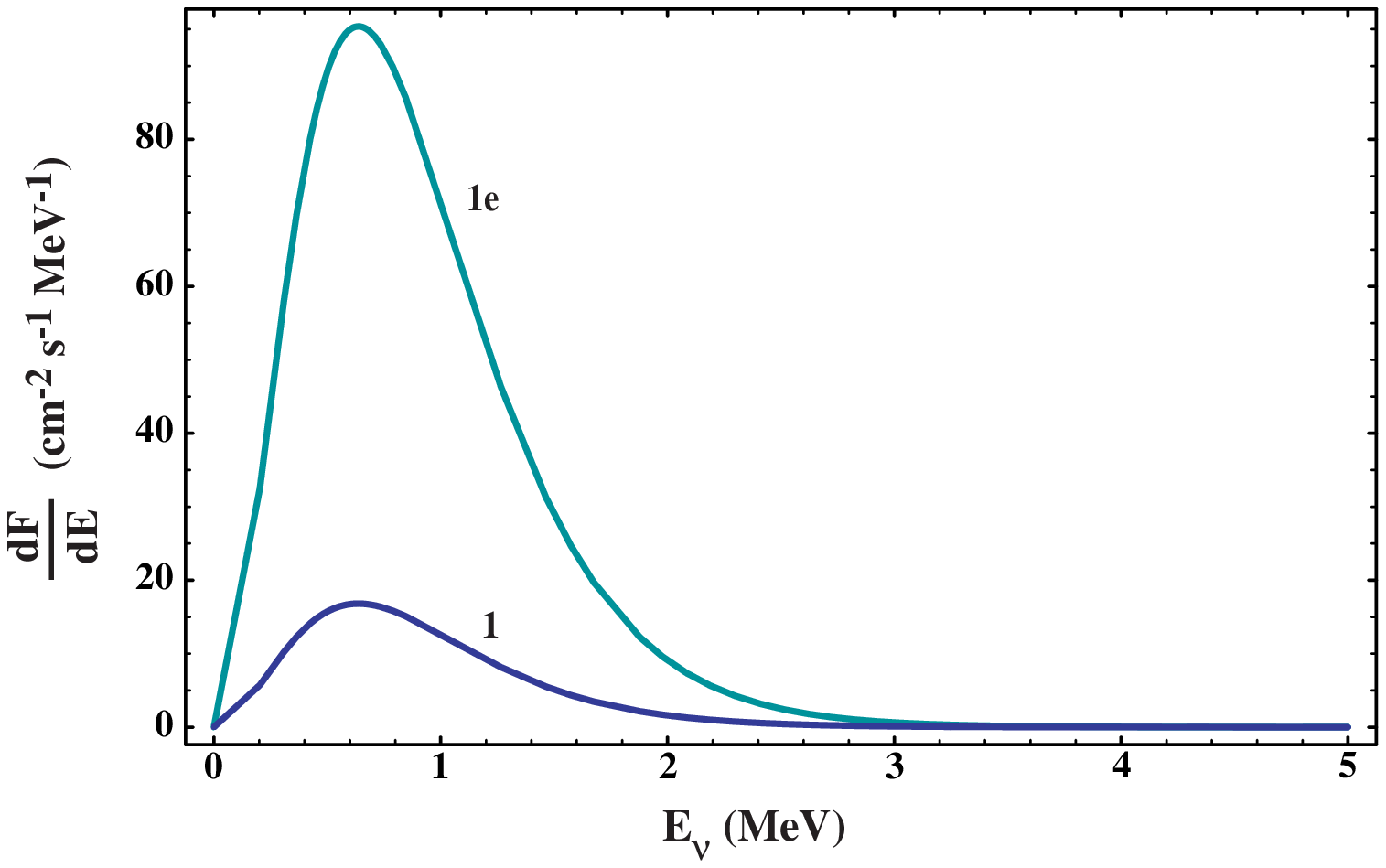}
\caption{Fluxes from the massive modes of Model 2.1 (blue) and 2.1e (cyan).}
\label{model1s2}
\end{figure}

The massive modes of Model 2.1 and 2.1e fluxes are plotted in Figure~\ref{model1s2} for the specific choice of $M_\mathrm{min} = 10^7$ M$_\odot$ which is the preferred case in \cite{daigne2}. As seen in Figure~\ref{sfr}, massive stars associated with Population III turn on
at a redshift of approximately 16, but the duration of the burst is relatively brief.
As a result, the peak of the flux distribution is at relatively low energy. More importantly,
because of the brevity of the burst, the {\em entire} neutrino spectrum is redshifted down,
in contrast to the Model 2.0 spectrum which extends to higher energy due to stars produced at lower
redshifts.
As expected, the more extreme model, 2.1e, has a peak flux which is about 5 times that found for Model 2.1.
This is directly related to the increased SFR in Model 2.1e as characterized by the increase in $\nu_2$.

\begin{figure}[ht]
\centering
\includegraphics[width=0.58\textwidth]{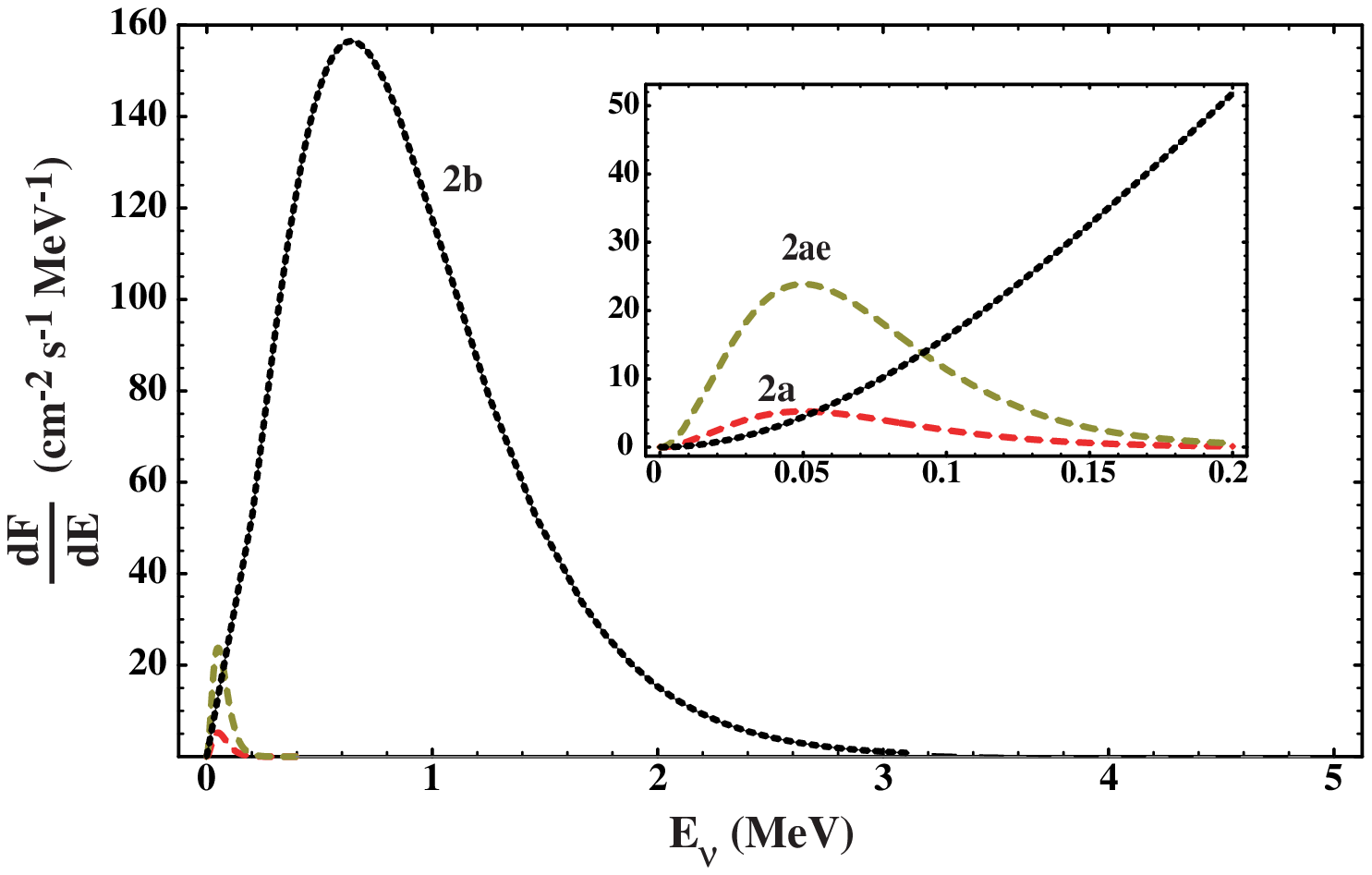}
\caption{Fluxes from the massive modes of Model 2.2a (dashed red), 2.2ae (dashed brown), and 2.2b (dotted black).
\label{model2abs2}}
\end{figure}

\begin{figure}[ht]
\centering
\includegraphics[width=0.58\textwidth]{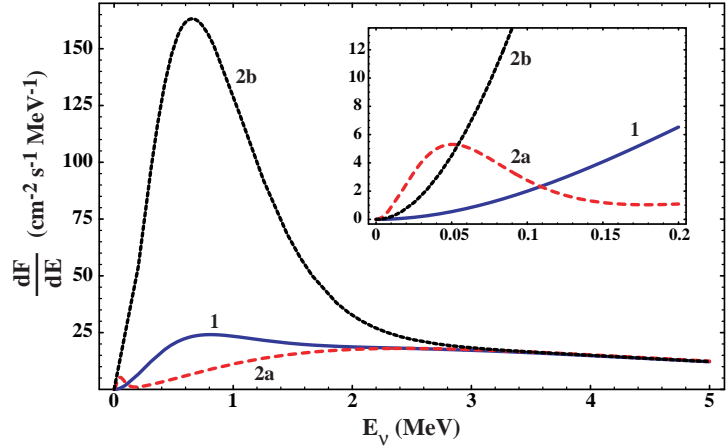}
\caption{Total fluxes for Model 2.1(solid blue), 2.2a (dashed red), and 2.2b (dotted black).
\label{totalfluxs2}}
\end{figure}

\begin{figure}[ht]
\centering
\includegraphics[width=0.58\textwidth]{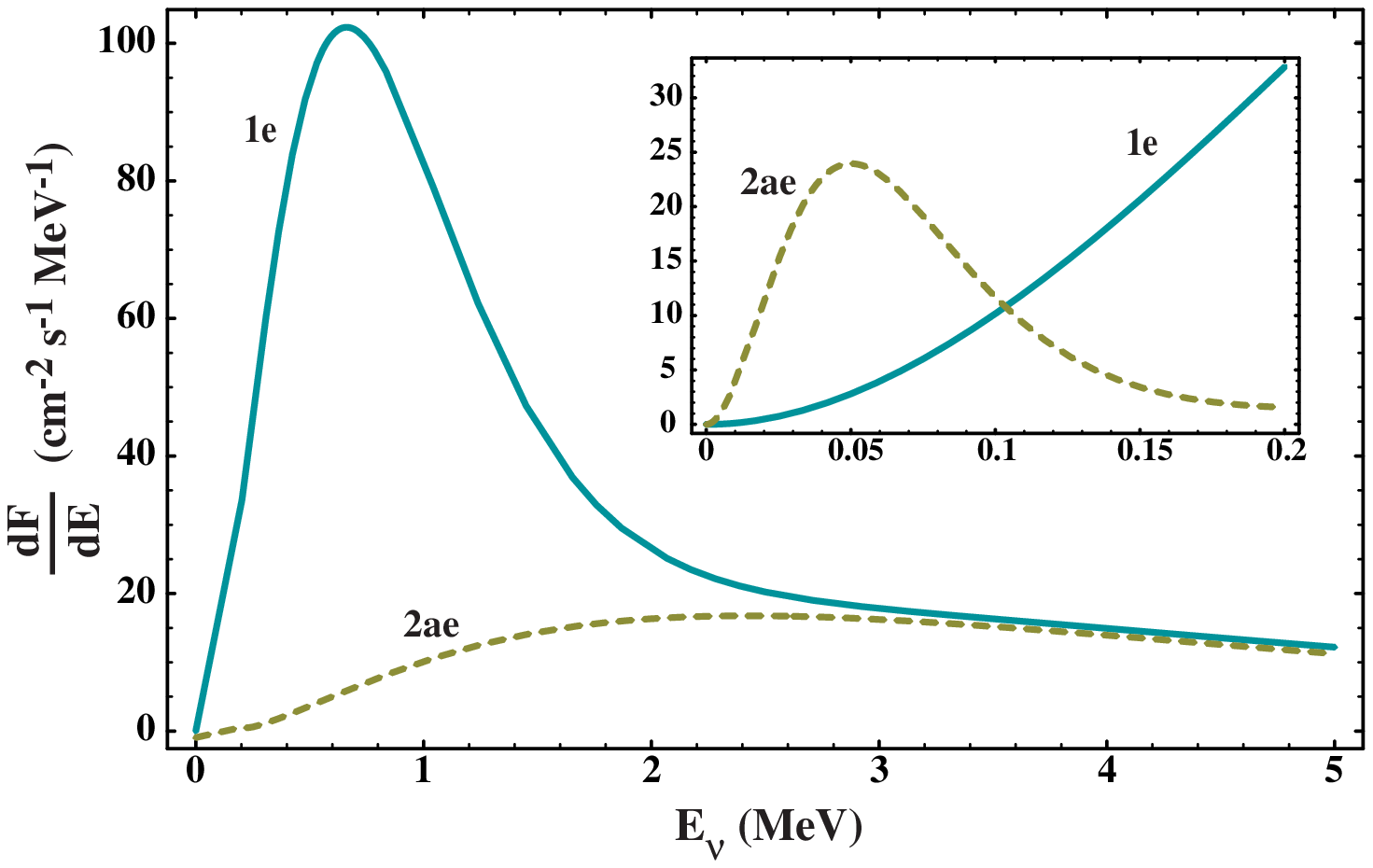}
\caption{Total fluxes  for Model 2.1e (solid cyan) and 2.2ae (dashed brown).
\label{totalfluxes2}}
\end{figure}

Similarly, we show in Figure~\ref{model2abs2} the resulting flux from very massive Population III stars
corresponding to Models 2.2a, 2.2ae and 2.2b.  As before, the fluxes from Models 2.2a are relatively small
and peak at very low energy as seen in the insert to the figure.
In Figures~\ref{totalfluxs2}  (\ref{totalfluxes2}), we show the total fluxes in Models 2.1, 2.2a, and 2.2b
(2.1e and 2.2ae) with $M_\mathrm{min} = 10^7$~M$_\odot$.  
As one expects, the low energy spectrum is 
dominated by neutrinos produced in the massive mode, whereas the spectrum at
higher energies ($E_\nu \ga 3$ MeV), is indistinguishable between the models and dominated
by the normal mode.

\subsubsection{Comparisons to other models}\label{sec:results:SRNbg:comp}

The differences in the fluxes between stages 1 and 2 models can be attributed entirely to differing IMF slopes, the baryon fraction,  and treatment of the SFR. In order to compare our calculations to previous ones, it is necessary to single out individual models and discuss the differences. 

Our Model 0 can be compared to models described in \cite{tot,ando,as,sksw}. 
In \cite{tot}, the calculated neutrino flux peaks at  
$ \simeq 8~\textrm{cm}^{-2}\textrm{s}^{-1}\textrm{MeV}^{-1}$ with neutrino 
energies of $\sim 3\,\textrm{MeV}$. In the models considered, star formation only occurs for $z\leq5$ and the average neutrino energy at emission was assumed to be a step function in which energies are generally a few MeV larger than our adopted value of $\langle E_{\bar{\nu}_e}\rangle = 15.3 \,\textrm{MeV}$. Since the flux changes as 
\beq
\frac{dF}{dE} \propto \frac{1}{{\langle E_{\nu} \rangle}^4} \frac{1}{e^{x/{\langle E_{\nu}\rangle}}+1}
\eeq
where $x$ depends on $z$, this both reduces the peak height and shifts the peak to larger neutrino energy at detection. 
In addition,  they include only supernovae that collapse to neutron stars, with 8 M$_{\odot} \leq m \leq 50$ M$_{\odot}$. In the absence of any component of stars which collapse to black holes, the flux should be smaller than that obtained by our calculations.
Ando, Sato and Totani \cite{ando} use a crude supernova model, but provide a more careful treatment of oscillations. As will be shown in section~\ref{sec:osc}, our results are qualitatively similar despite the fact that our integrated flux differs from theirs by an order of magnitude.

The more modern approach taken in \cite{ando,as} is based on the cosmic star formation rate.
However, our Model 2.0 spectra have significantly higher peaks than those in \cite{ando,as}. First,  
improved determinations of the SFR at high redshift are higher by a factor of about 2.   A more significant difference between our results can be traced to our inclusion of stars that collapse to black holes.  The change in gravitational energy is much greater when the remnant is a black hole than when it is a neutron star leading to a larger value for $E_{cc}$, and therefore the neutrino luminosity is also much greater.  For comparison, if we consider only stars that collapse to neutron stars, our flux is reduced by a factor of $\sim 5$.

In \cite{sksw}, the SFR is parametrized as a broken power law, flat for $z>1$ and fit to the observed cosmic star formation rate.  As in \cite{tot,ando}, star formation begins at $z=5$. The effect of neglecting star formation at higher redshift is inconsequential for the higher energy tail of the flux, but both diminishes the lower energy flux and shifts the peak forward. Both of these effects are consequences of the fact that neutrinos from the earliest supernova events will arrive at Earth with the lowest energies as a result of redshift.

The flux from a massive mode of star formation was computed in \cite{immrs}. Population III stars were assumed to be rotating 300 M$_{\odot}$ stars with a SFR approximated by a delta function peaked at $z=17$. We observe the same peak location and spectral shape in our Models 1 and 2b.  
Our fluxes peak at larger values than theirs for two main reasons. First, our integrated fraction of baryonic matter in Pop III stars is larger. They use a fixed value of $10^{-3}$, whereas our value is derived for each model; $2.2 \times 10^{-3}$ for Model 2.1 and $7.3 \times 10^{-3}$ for Model 2.2b. Second, in Model 2.2b the stars collapse entirely to black holes, while in their model the energy emitted in neutrinos corresponds roughly to the gravitational energy released when the stars collapse to black holes with masses equivalent to the mass of the helium core before collapse. As a result, a 300\,M$_{\odot}$ star in our model emits $\sim 2.7$ times the number of neutrinos as the same star in their model.

\subsection{Thermonuclear neutrinos}

In this section we consider briefly the neutrinos emitted by stars during the hydrogen-burning phase. 
The CNO cycle is the dominant neutrino production mechanism, responsible for $90\%$ of the thermonuclear neutrino flux. The remaining $10\%$ is due to the pp chains.  To simplify the calculations, we consider only the reactions given in Table \ref{therm} below.

\begin{table}[ht]
\begin{center}
\begin{tabular}{|l|c|c|}
\hline
Reaction & Endpoint & \% \\
\hline
p$+$p$\to e^{+}+\nu_e$ & 0.42 & 10 \\
${}^{13}\textrm{N} \to {}^{13}\textrm{C}+e^{+}+\nu_e $ & 1.20 & 45  \\
${}^{15}\textrm{O} \to {}^{15}\textrm{N}+e^{+}+\nu_e $ & 1.73 & 45  \\
\hline
\end{tabular} 
\caption{These are the thermonuclear reactions considered here.  The endpoint is indicated in MeV, and the percentage refers to the amount of the total thermonuclear neutrino flux due to each reaction mechanism.}
\label{therm}
\end{center}
\end{table}

\begin{figure}[ht]
\centering
\includegraphics[width=0.58\textwidth]{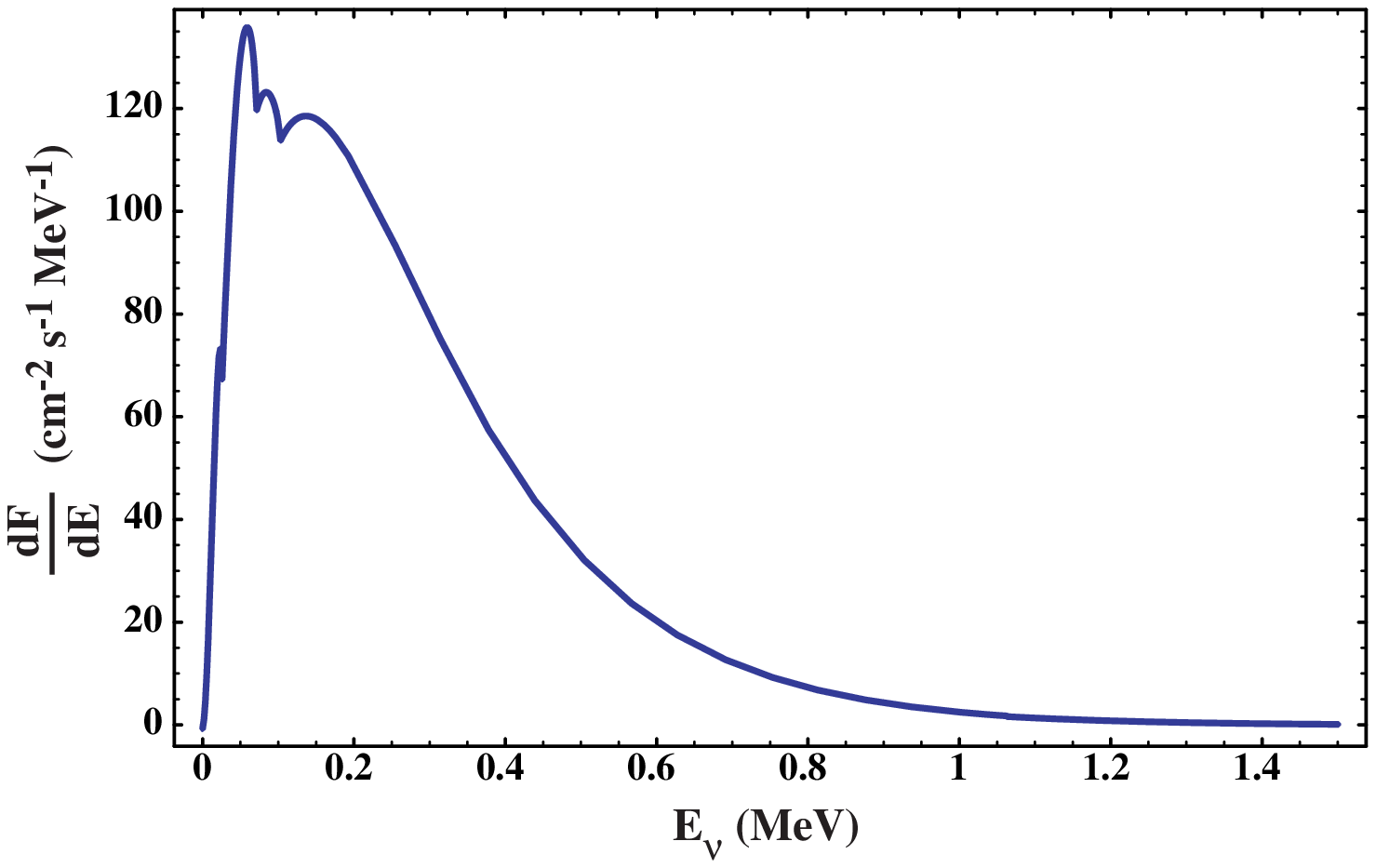}
\caption{Total flux of thermonuclear neutrinos from Model 2.1.
\label{therm71}}
\end{figure}

The total number of neutrinos emitted by a star of mass $m$ during hydrogen burning has been estimated to be~\cite{immrs}
\begin{equation}
N_{therm-nucl} \approx 0.2 \frac{m}{m_N},
\end{equation}
where $m_N$ is the mass of a nucleon, and the energy spectrum at emission is taken from Ref.~\cite{bahcall}. The total flux of thermonuclear electron neutrinos in  Model 2.1 (including the normal  
mode) is shown in Figure~\ref{therm71}. As one can see, the total flux is quite large and significant up to $E_{\nu_e} \approx 1\,$MeV. The flux from the massive component is cut off at $E_{\nu_e} \approx 0.1\,$MeV, which is far below the neutrino energy threshold at gallium experiments such as SAGE and Gallex-GNO ($E_{threshold} \approx 0.233$MeV~\cite{bahcall}). The normal mode thermonuclear flux is smooth and broad due to the larger SFR at low redshift. However, the flux is overshadowed by several orders of magnitude by solar neutrinos from the $pp$ chains and the CNO cycle. Since experiments sensitive to this energy range, like Borexino and LENS, do not have the capability to resolve any directional information about incoming neutrinos, the only current possibility for distinguishing SRN's from solar neutrinos is by spectral shape.

\section{Detection}\label{sec:det}

The detection of SRN's is inhibited mainly by difficulties excluding background events which include solar neutrinos, atmospheric neutrinos and antineutrinos, and antineutrinos from nuclear reactors. Cosmic ray muons also produce events that mimic the expected signal.

The solar neutrino flux at Earth is larger than the expected flux from SRN's by several orders of magnitude for $E_{\nu} \lesssim 19\,$MeV~\cite{as}. However, neutrinos, rather than antineutrinos, are produced in the thermonuclear reactions in the sun and have a smaller cross section for detection by about two orders of magnitude. This, with the directional information from recoil electrons in the detector, allows this background to be excluded at SK, KamLAND, and SNO. But decays of spalled nuclei from cosmic ray muons constitute an unavoidable background in this range.

Atmospheric electron antineutrinos are cause for concern above $\sim8\,$MeV, but the flux becomes larger than the expected flux of SRN's only for $E_{\nu}\gtrsim35\,$MeV~\cite{sksw}. An atmospheric muon neutrino can interact with a nucleus to form a muon, which will be invisible in Cerenkov detectors if its kinetic energy is below the Cerenkov radiation threshold of 53 MeV.  This background is significant for $19\,$MeV$<E_{\nu}<35\,$MeV, but it can be described by the Michel spectrum and was subtracted off to obtain the current upper limit for the flux of SRN's at SK of $1.2\,$cm$^{-1}$s$^{-1}$ for $E_{\nu}>19.3\,$MeV~\cite{SKlimit}.

It has been pointed out recently \cite{sno}  that if the background analysis from SK is coupled with the sensitivity to electron neutrinos at SNO it will be possible to reduce the upper limit on the flux of electron neutrinos. SNO should be sensitive to a flux of $6\,$cm$^{-2}$s$^{-1}$ in the range $22.5\,$MeV$<E_{\nu_e}<32.5\,$MeV, which is an improvement on the Mont Blanc limit by three orders of magnitude. 

\begin{figure}[ht]
\centering
\includegraphics[width=0.58\textwidth]{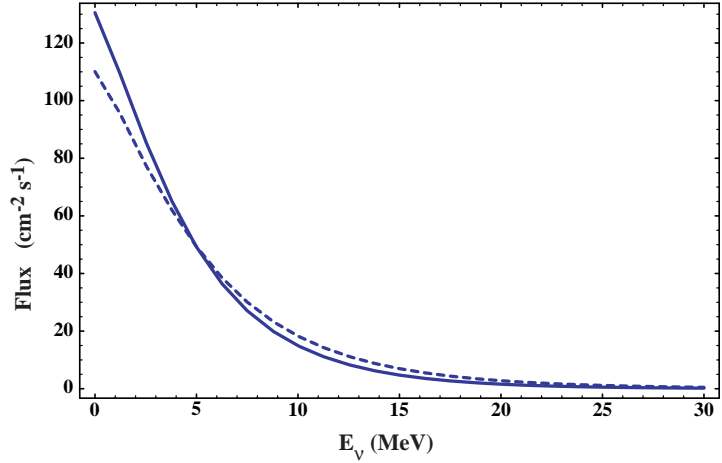}
\caption{Detectable fluxes from Model 2.1 with (dashed blue) and without (solid blue) oscillations as a function of neutrino energy threshold.\label{threshold}}
\end{figure}

In Figure~\ref{threshold}, we show the observable flux 
\beq
F(E_\mathrm{thresh}) = \int^\infty_{E_\mathrm{thresh}} {dF \over dE} dE
\eeq
as a function of detector threshold energy. 
While the fluxes are quite appreciable at low threshold energies, they in fact remain
relatively high at larger energies due to the large SFR associated with Model 0.
Indeed, in all of our stage 2 models, our predicted flux above 19.3 MeV already exceeds
the current bound of  $1.2$ cm$^{-1}$s$^{-1}$ from SuperK~\cite{SKlimit}. The detailed flux
predictions are given in 
Table~\ref{numberflux}, where we show the detectable flux for the viable energy windows at SK and SNO. Although SRN's will likely not be seen at SNO given these flux levels, in many of our models the SK bound is saturated by the expected flux, indicating that SRN's may be observed in the near future.
This is in agreement with previous arguments made in \cite{sksw} based on simplified evolution models
as well as arguments based on SN1987A \cite{lun}.

\begin{table}[ht]
\begin{center}
\begin{tabular}{|l|c|c|}
\hline
Model & SK Flux & SNO Flux \\
\hline
\hline
1.0 & 0.40 & 0.11 \\
1.1 & 0.40 & 0.11 \\
1.2a & 0.40 & 0.11 \\
1.2b & 0.40 & 0.11 \\
\hline
2.0 & 1.8 & 0.47 \\
2.1 & 1.8 & 0.47 \\
2.1osc & 3.2 & 1.4 \\
2.1e & 1.9 & 0.49 \\
2.2a & 1.9 & 0.48 \\
2.2ae & 1.9 & 0.49 \\
2.2b & 1.8 & 0.47 \\
2.2bosc & 3.2 & 1.4 \\
\hline
\end{tabular} 
\caption{Predicted fluxes in cm$^{-2}$s$^{-1}$ in the models considered here. Results are given for electron antineutrinos with energies $E_{\bar{\nu}_e}>19.3$MeV for SK and for electron neutrinos with $22.5$MeV$<E_{\nu_e}<32.5$MeV for SNO. \label{numberflux}}
\end{center}
\end{table}

Despite the large fluxes displayed in Table \ref{numberflux} relative to the SK limit \cite{SKlimit}, 
one can not conclude that the stage 2 models considered have already been excluded by experiment.
There are of course many uncertainties built into our chemical evolution models as well as uncertainties
in the adopted neutrino physics.  For example, one of the differences between our stage 1 and stage 2 models, is our choice of the IMF.  In stage 1 models, the slope of the IMF was fixed at 1.7 whereas
in stage 2 models, it is fixed at 1.3.  The impact of this difference lies in the strong suppression of
massive stars, i.e. the precursors of neutrino producing supernovae. As one can see from the table,
the steeper IMF sufficiently suppresses the neutrino flux to satisfy the SK bound.

\begin{figure}[ht]
\centering
\includegraphics[width=0.58\textwidth]{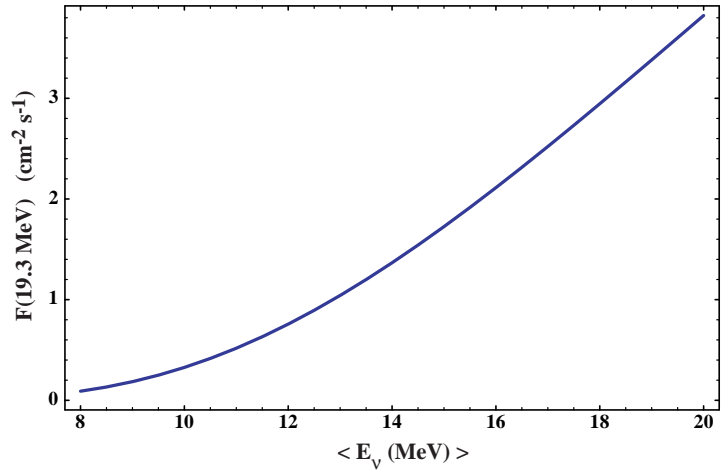}
\caption{The integrated flux above 19.3 MeV in Model 2.1 as a function of the
average neutrino energy.
\label{aveE}}
\end{figure}

One can also see from Table \ref{numberflux} that the dominant contribution to the integrated
flux above the SK threshold comes from normal mode stars. Even the more extreme model
2.1e only contributes 0.1 cm$^{-2}$ s$^{-1}$ to the flux above 19.3 MeV.
However, these fluxes are very sensitive to our assumed average neutrino energy.
Recall our adopted value for $E_{\bar \nu_e}$ is 15.3~MeV.  In Fig. \ref{aveE}, we show the sensitivity
of the flux above 19.3 MeV ($F(19.3)$) to the average neutrino energy.
In order to satisfy the SK limit of 1.2 cm$^{-2}$s$^{-1}$, we would have to lower
$\langle E_{\bar \nu_e} \rangle$ to 13.3 MeV.  This is fully consistent with the range of neutrino
energies in supernova models \cite{krj}.

Ill-understood backgrounds and large detector energy thresholds make detection of SRN's difficult. Because the SRN flux peaks at or below $1\,$MeV, there is little hope that the peak will be probed with existing experiments. However, being able to exclude backgrounds at lower energies would greatly increase the probability for observation at SK or a similar experiment.

\section{Effects of Oscillations}\label{sec:osc}

Due to our assumption of equipartition of energy among the neutrino species, oscillation will only affect the flux if different species emerge from the explosion with different average energies. 
Indeed most supernova model calculations do show a hierarchy of neutrino energies, and 
because the energies of $E_{\bar \nu_\mu}$ and $E_{\bar \nu_\tau}$ are generally higher
than $E_{\bar \nu_e}$, the effect of oscillations will in general increase the observable flux.
The effects of oscillations on the neutrino background were considered previously in \cite{osc}.

\begin{figure}[ht]
\centering
\includegraphics[width=0.58\textwidth]{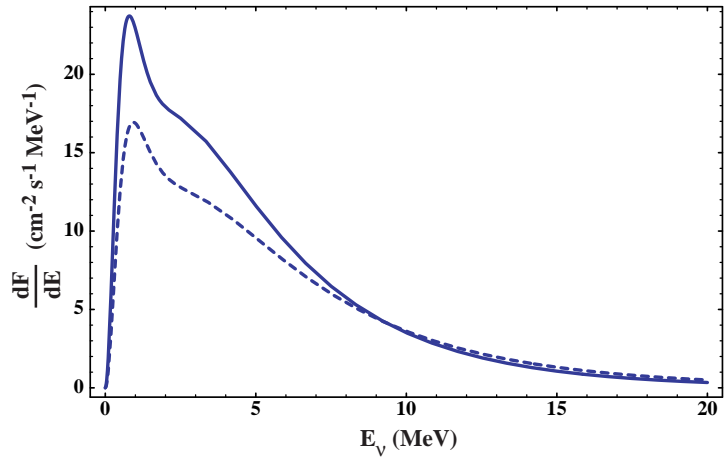}
\caption{Total fluxes of anti-electron neutrios from Model 2.1 with (dashed blue) and without (solid blue) oscillations.
\label{osc}}
\end{figure}

Figure~\ref{osc} shows the total flux of electron antineutrinos from Model 2.1 both with and without oscillations, where we use the neutrino average energies discussed in section~\ref{sec:SRN} and maximal mixing has been assumed. Although the total number of electron antineutrinos arriving at Earth is smaller due to oscillations by $\sim 16\%$, the flux of neutrinos with energies greater than $\sim 9.5\,$MeV is larger. The differential flux for Model 2.1 with and without oscillations is shown in Fig. \ref{osc}. 
Note that the solid curve here is identical to that in Fig. \ref{totalfluxs2}. Here we clearly see
the individual peaks due to the massive mode at $\la 1$ MeV and normal mode at $\sim 3$ MeV. 

For $E_\nu >19.3\,$MeV, oscillation effects would increase the observable flux at SK by as much as $78\%$. This effect is seen in Fig. \ref{threshold} where one sees that the integrated flux with oscillations (dashed curve) exceeds the flux when oscillations are ignored for threshold energies greater than about 6 MeV. The effect of oscillations on the SK observable flux is seen in Table \ref{numberflux} for the models labeled 2.1osc and 2.2bosc.  Reconciling these fluxes with the SK limit would require a further 
drop in the average neutrino energy or a tightening of the assumed neutrino energy hierarchy.
Similarly, the flux of electron neutrinos potentially observable at SNO is increased by almost $200\%$ in the energy window $22.5$MeV$<E<32.5$MeV. Although our flux calculated with oscillations is still less than 1/4 that necessary to approach the projected SNO sensitivity~\cite{sno}, with a better understanding of backgrounds the prospects for detection in the near future are encouraging.

\section{Conclusions}\label{sec:conc}

We have considered several scenarios for star formation which reproduce the observed chemical abundances and SFR for $z \leq 6$ and reionize the universe at high redshift. Each model of star formation here consists of a normal mode coupled to a  Population III mode of massive star formation at high redshift. We examined the SRN flux from the core collapse supernova explosion as well as from the thermonuclear burning stage. 

Because the massive mode of star formation is so brief and takes place at high redshift, the corresponding electron anti-neutrino fluxes peak at $E_{\nu} \lesssim 1\,$MeV. 
Thus despite the large fluxes produced by the massive mode, these low energy neutrinos
will be difficult to detect. In contrast, 
the normal mode of star formation, which dominates the flux at observable energies, is peaked at a somewhat higher energy and has a broad spectrum due to the production of stars at lower redshift. The neutrinos produced during thermonuclear burning are emitted with much lower energies, on average, but also exhibit the behavior of a sharp peak due to the massive mode added to a broader spectrum from the normal mode. Thermonuclear neutrinos are unlikely to be observed at experiments which cannot at least partially resolve the direction of the incoming particle because they are produced in the electron flavor state and the spectrum lies several orders of magnitude below that of solar neutrinos. Our calculated fluxes of SRN's from core collapse, however, saturate the SK bound of $1.2\,$cm$^{-2}$s\,$^{-1}$ for $E_{\nu}>19.3\,$MeV in all stage 2 models. Although there are uncertainties in the neutrino physics, such as the average energies at emission, the prospects for observation in the near future are good. We also examined the effect of oscillations by calculating the flux with maximal mixing. With the accepted neutrino average energy hierarchy, $\langle E_{\nu_e}\rangle < \langle E_{\bar{\nu}_e}\rangle < \langle E_{\nu_x}\rangle$, any oscillation will harden the high energy tail of both the $\nu_e$ and the $\bar{\nu}_e$ spectra.

Further refinement of the neutrino physics and measurement of the SFR out to higher redshift would allow for a more definite flux prediction. With decreased detector thresholds and increased background rejection, observation of the SRN flux will soon be possible.

{\bf Acknowledgements}

The work of K.A.O., F.D. and E.V. was supported by the Project "INSU - CNRS/USA", and the work of K.A.O. and P.S. was also supported partly by DOE grant
DE--FG02--94ER--40823.

\end{document}